\documentclass[conference]{IEEEtran}
\IEEEoverridecommandlockouts
\usepackage{cite}
\usepackage{amsmath,amssymb,amsfonts}
\usepackage{algorithmic}
\usepackage{graphicx}
\usepackage{textcomp}
\usepackage{xcolor}
\usepackage{braket}
\usepackage{caption}
\usepackage{subcaption}
\usepackage[ruled,vlined]{algorithm2e}
\usepackage{multicol}
\usepackage{widetext}
\def\BibTeX{{\rm B\kern-.05em{\sc i\kern-.025em b}\kern-.08em
    T\kern-.1667em\lower.7ex\hbox{E}\kern-.125emX}}
\begin{document}

\title{Discriminating Quantum States with Quantum Machine Learning\\
\thanks{Funding: Department of Energy, ASCR Quantum Testbed Pathfinder program}
}

\author{\IEEEauthorblockN{David Quiroga}
\IEEEauthorblockA{\textit{Engineering Faculty} \\
\textit{Universidad de Antioquia}\\
Medellín, Colombia \\
andres.quiroga@udea.edu.co}
\and
\IEEEauthorblockN{Prasanna Date}
\IEEEauthorblockA{\textit{Computer Science and Mathematics} \\
\textit{Oak Ridge National Laboratory}\\
Oak Ridge, United States \\
datepa@ornl.gov}
\and
\IEEEauthorblockN{Raphael Pooser}
\IEEEauthorblockA{\textit{Computational Science and Engineering} \\
\textit{Oak Ridge National Laboratory}\\
Oak Ridge, United States \\
pooserrc@ornl.gov}
}

\maketitle

\begin{abstract}
Quantum machine learning (QML) algorithms have obtained great relevance in the machine learning (ML) field due to the promise of quantum speedups when performing basic linear algebra subroutines (BLAS), a fundamental element in most ML algorithms. By making use of BLAS operations, we propose, implement and analyze a quantum k-means (qk-means) algorithm with a low time complexity of \(\mathcal{O}(NKlog(D)I/C)\) to apply it to the fundamental problem of discriminating quantum states at readout. Discriminating quantum states allows the identification of quantum states \(\ket{0}\) and \(\ket{1}\) from low-level in-phase and quadrature signal (IQ) data, and can be done using custom ML models. In order to reduce dependency on a classical computer, we use the qk-means to perform state discrimination on the IBMQ Bogota device and managed to find assignment fidelities of up to 98.7\% that were only marginally lower than that of the k-means algorithm. Inspection of assignment fidelity scores resulting from applying both algorithms to a combination of quantum states showed concordance to our correlation analysis using Pearson Correlation coefficients, where evidence shows cross-talk in the (1, 2) and (2, 3) neighboring qubit couples for the analyzed device.
\end{abstract}


\section{Introduction}
Quantum Computing represents a powerful computation model able to efficiently perform basic linear algebra subroutines (BLAS) through the use of unitary matrices that make rotations on qubits to control their state. These quantum BLAS or qBLAS enable matrix operations on vectors in a high-dimensional vector space, and many machine learning protocols are based on such operations \cite{Biamonte2017}. Machine learning algorithms have benefitted from quantum computers as speedups have been demonstrated through the use of qBLAS \cite{PhysRevLett.103.150502}\cite{PhysRevLett.109.050505}\cite{Childs_2017}. Despite the potential quantum computers have in machine learning, the noise present in noisy intermediate-scale quantum (NISQ) computers has proven to be a challenging problem \cite{Preskill2018quantumcomputingin}. Coherent and incoherent errors \cite{Feng_2016} interact with quantum computers in the form of pulse shaping errors \cite{PhysRevResearch.2.043418}, readout errors \cite{Nachman2020}, cross-talk \cite{10.1145/3373376.3378477}\cite{8614500}\cite{Sarovar2020detectingcrosstalk}, dephasing \cite{HU_2002} and others, leading to lower gate fidelity due to low coherence times and leakage into higher energy states \cite{werninghaus2020leakage}\cite{PhysRevA.93.060302}\cite{Chasseur_2015} and eventually producing less accurate quantum algorithms. Pulse and gate-level benchmarks such as Randomized Benchmarking \cite{Knill_2008}, the Benewop benchmark \cite{PhysRevA.101.042308} and Quantum Process Tomography \cite{10.5555/1972505} focus on different types of errors to measure them and provide information on how noise sources impact a quantum computer's performance.

Our research provides an analysis of errors occurring at readout that hint towards cross-talk in IBMQ's Bogota device by using Pearson Correlation coefficients and state discrimination with two algorithms. State discrimination is one of the internal processes a quantum computer undergoes at readout that enable telling one state apart from another based on low-level signals retrieved after applying a quantum gate to a qubit. In state discrimination, machine learning models are fitted to signal data and tasked with classifying each data point to states \(\ket{0}\) and \(\ket{1}\). Although the current way of performing state discrimination in IBM quantum computers is by applying a Linear Discriminant Analysis to data, it is also possible to use other machine learning algorithms \cite{Magesan_2015}\cite{Alexander_2020}. We propose a quantum machine learning model based on the k-means algorithm with lower computational complexity and apply it in state discrimination at readout in order to reduce dependency on a classical computer with a potential to require less resources and time in the future. Reduction of state preparation and measurement (SPAM) errors may also improve the performance of the quantum k-means, as well as most quantum machine learning algorithms. On the other hand, we use Pearson Correlation coefficients to provide information on correlations occurring between sets of signal data. Cross-talk can thus be determined by using both perspectives in conjunction when analyzing pairs of neighboring qubits on a quantum device.

The rest of the paper is structured as follows. In section II, we mention related work attending state discrimination and quantum machine learning. In section III, we describe our implementation of the qk-means algorithm and provide a complexity analysis. In section IV, we explain the methods used for our research. In section V, we provide results on cross-talk and the performance of the k-means and the qk-means algorithm and discuss them. In section VI, we give our conclusions on our work and comment on future work for our research.

\section{Related work}

\subsection{Discriminating quantum states}

ML techniques have been used for the discrimination of quantum measurement data for some time now as an approach to retrieve amplitudes resulting from execution of a quantum circuit. One application of these techniques on quantum measurement trajectories can be seen in \cite{Magesan_2015}, where SVM, k-means, LDA, and the RUSBoost algorithms are used to discriminate data belonging to the \(\ket{0}\) and the \(\ket{1}\) states. While these discriminators are not offered out-of-the-box, Qiskit Pulse enables the use of different ML models to work as readout discriminators and also offers a Linear Discriminant Analysis module to use as a custom discriminator \cite{Alexander_2020}. The main difference in the studies performed on quantum measurement trajectories with the way quantum states are being discriminated in Qiskit Pulse is the form of the data, given that in-phase and quadrature (IQ) signal data is returned from computations on IBMQ devices that recieve schedules as inputs.

On IBM quantum computers, the task of state discrimination corresponds to the first and second level of their quantum hardware's obtainable results through OpenPulse \cite{mckay2018qiskit}, mapping the phase (IQ) signals of their first levels onto qubit states and counts. While state discrimination can be performed both on level 0 and on level 1, they solve the problem through a different type of data input. Level 0 focuses on classifying shot trajectories through Machine Learning by using filter functions for each trajectory \cite{doi:10.1063/1.4813269} and algorithms such as Quadratic Discriminant Analysis (QDA) and Support Vector Machines (SVM). ML algorithms for discrimination may be specified on Qiskit if they are formatted as a scikit-learn classifier \cite{sklearn_api}. While QML algorithms have not been used to perform discrimination at readout, it is possible to achieve this by specifying QML algorithms as scikit-learn classifiers, which is the solution we implement in this paper.

\subsection{Quantum Machine Learning}

QML focuses on the use of quantum algorithms, which harness the capabilities of quantum mechanics, to solve common ML problems and to improve existing algorithms \cite{Schuld_2014,adachi2015application}. Many algorithms implement quantum basic linear algebra subroutines (qBLAS) that have shown exponential quantum speedups over the most optimal methods available in classical computation \cite{Biamonte2017}. The use of amplitude amplification \cite{Brassard_2002} and the HHL \cite{PhysRevLett.103.150502} algorithm in QML algorithms results in speedups in computational and query complexity over their classical counterparts. Methods such as Bayesian Interference \cite{PhysRevA.89.062315}, PCA \cite{Lloyd2014}, Support Vector Machines \cite{PhysRevLett.113.130503} and many others \cite{wiebe2016quantum}\cite{PhysRevLett.109.050505}\cite{PhysRevLett.117.130501} have adapted implementations of qBLAS to achieve speedups of up to $O(\sqrt{N})$ and $O(logN)$. Adiabatic quantum computers can also provide adequate solutions for ML problems provided that models can be represented as quadratic unconstrained binary optimization (QUBO) problems as is theorized in \cite{Date2021}, posing better time and space complexities than their classical counterparts. The balanced k-means clustering algorithm implemented on adiabatic quantum computers demonstrates performance similar to the best classical approaches \cite{arthur2020balanced}.
There exist several limitations on QML algorithms as only small-scale quantum computers and special-purpose quantum simulators are currently available to implement them. Particularly, feeding input by encoding classical data into quantum computers and obtaining the output from calculations performed constitutes some overhead \cite{Aaronson2015}, which can be alleviated to some extent using qRAM and operating directly on quantum data in a manner similar to variational methods \cite{Peruzzo2014}.

\section{Quantum K-Means}

The qk-means algorithm is an implementation of the k-means in which the distances between data points and cluster centroids are calculated using destructive interference on a quantum circuit. To achieve this, classical data must be mapped onto a quantum circuit, while a quantum algorithm for distance computation must be implemented. Our algorithm uses two data encoding strategies and sends jobs in batches to IBMQ devices. The algorithm was written in Python using the Qiskit library and several data preprocessing functions available through the scikit-learn library \cite{scikit-learn}. 

Out of the two encoding strategies, amplitude encoding of the input dataset onto qubits provides a convenient way to map \(n\)-dimensional input vectors, given the wide range of different values that can be obtained. Since qubit amplitudes must have a norm of 1, normalization of input vectors must be done in order to encode amplitudes onto quantum states.
That is, for an \(n\)-dimensional input vector 
$$
\mathbf{a} =\begin{pmatrix} a_0 & a_1 & ... & a_n \end{pmatrix}, \mathbf{a}^\prime = \frac{\mathbf{a}}{\left\lVert \mathbf{a}\right\rVert} = \begin{pmatrix} a^\prime_0 & a^\prime_1 & ... & a^\prime_n \end{pmatrix}
$$

will be encoded as 

$$
\ket{\psi} = a^\prime_0\ket{0} + a^\prime_1\ket{1} + ... + a^\prime_n\ket{n}.
$$

Another way of encoding data implemented by the qk-means algorithm is establishing a rotation angle \(\theta\) in which qubits are set with an \(Ry\) gate. As a specific case, given a \(2\)-dimensional input vector

$$
\mathbf{a} =\begin{pmatrix} a_0 & a_1 \end{pmatrix}, \theta = \arctan{\frac{a_1}{a_0}}
$$

is the angle that will be applied as an \(Ry(\theta)\) gate to a specific qubit.

\begin{figure}
\centering
\includegraphics[width=0.2\textwidth]{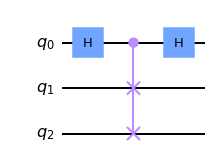}
\caption{SwapTest circuit for measuring distances between qubit 1 (second line) and qubit 2 (third line).}
\label{fig:SwapTest}
\end{figure}

The next important aspect is calculating distances using a quantum circuit. For this, a SwapTest circuit is applied as shown in Figure~\ref{fig:SwapTest} to find how much two quantum states differ and therefore act as a distance measurement in the form

$$
D(\ket{x},\ket{y}) = \sqrt{2-2|\braket{x|y}|}.
$$

\begin{figure*}
     \centering
     \begin{subfigure}[h]{0.45\textwidth}
         \centering
         \includegraphics[width=0.8\textwidth]{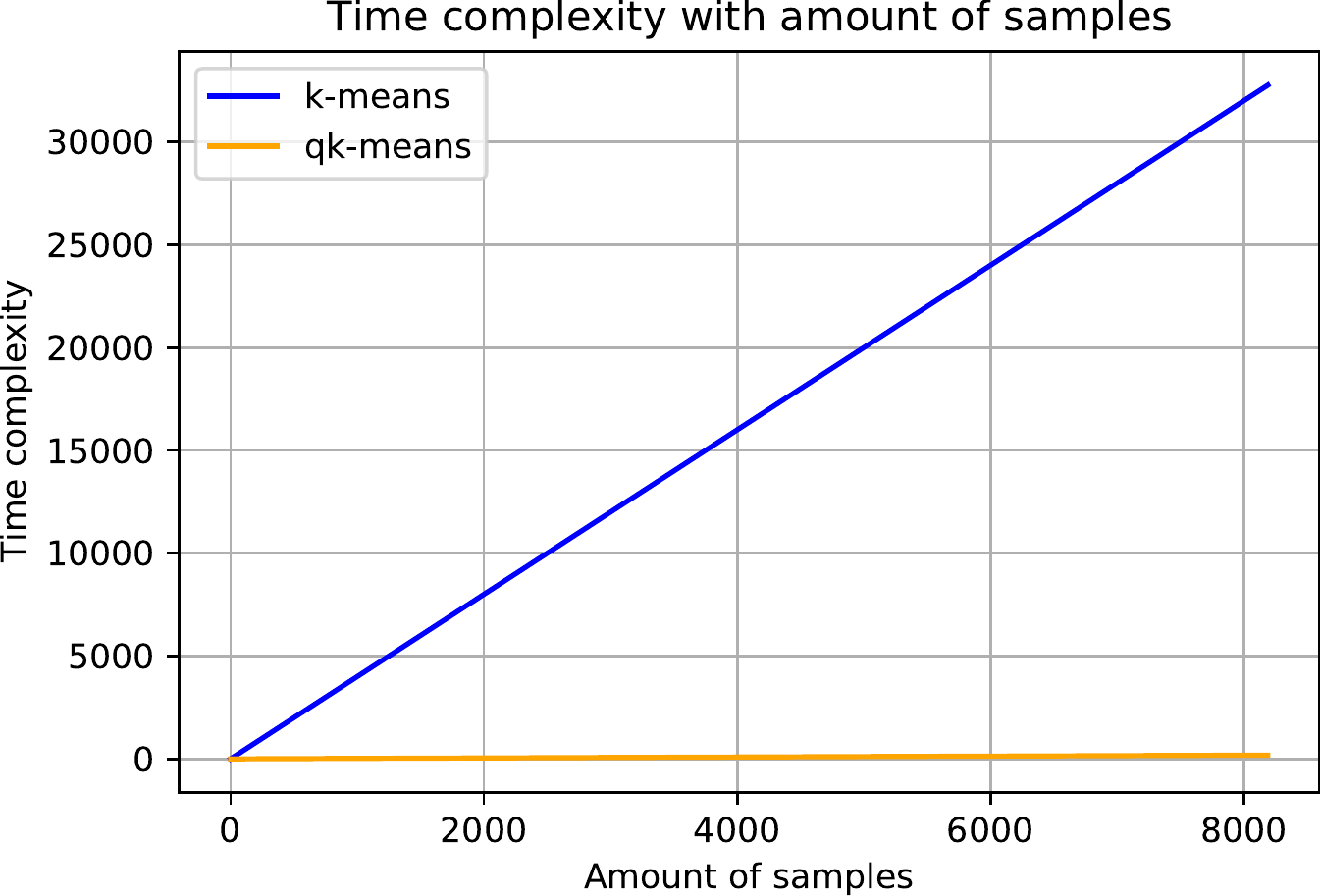}
         \caption{Classical and quantum time complexity plot depending on the amount of samples.}
         \label{fig:complexity_sample_q_k}
     \end{subfigure}
     \hfill
     \begin{subfigure}[h]{0.45\textwidth}
         \centering
         \includegraphics[width=0.8\textwidth]{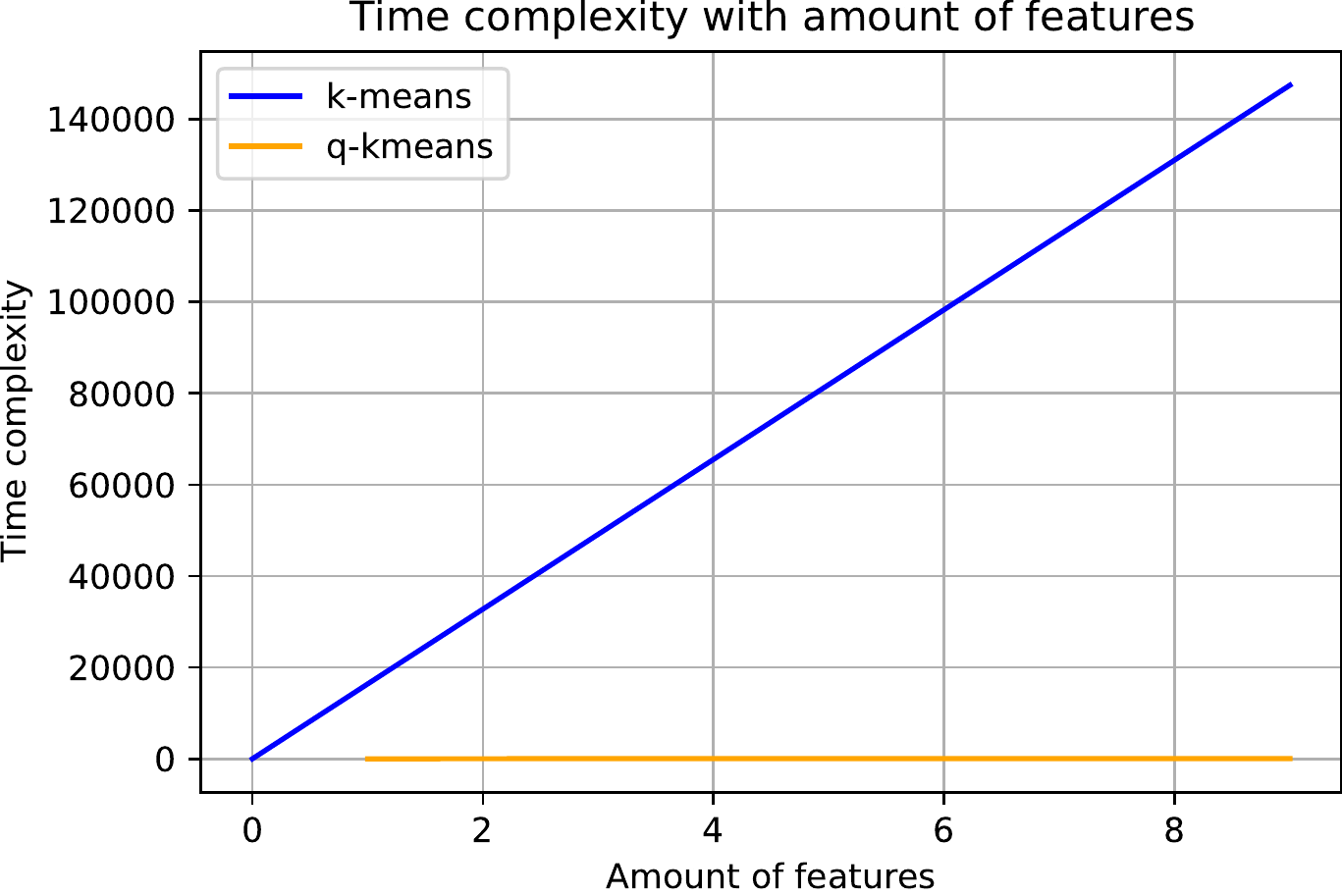}
         \caption{Classical and quantum time complexity plot depending on the amount of features.}
         \label{fig:complexity_feature_q_k}
     \end{subfigure}
     \hfill \\
     \begin{subfigure}[h]{0.45\textwidth}
         \centering
         \includegraphics[width=0.8\textwidth]{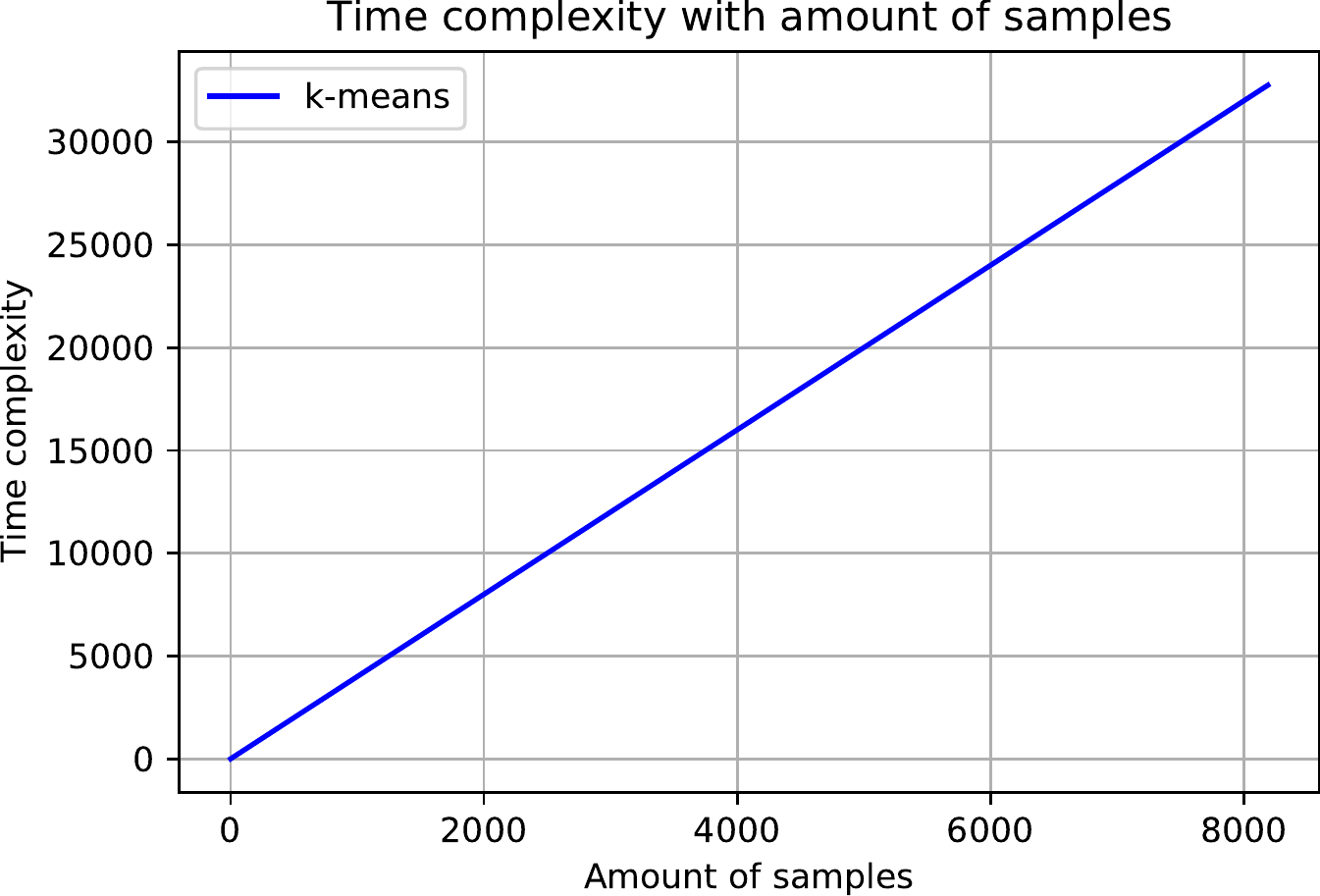}
         \caption{Classical time complexity plot depending on the amount of samples.}
         \label{fig:complexity_sample_k}
     \end{subfigure}
     \hfill
     \begin{subfigure}[h]{0.45\textwidth}
         \centering
         \includegraphics[width=0.8\textwidth]{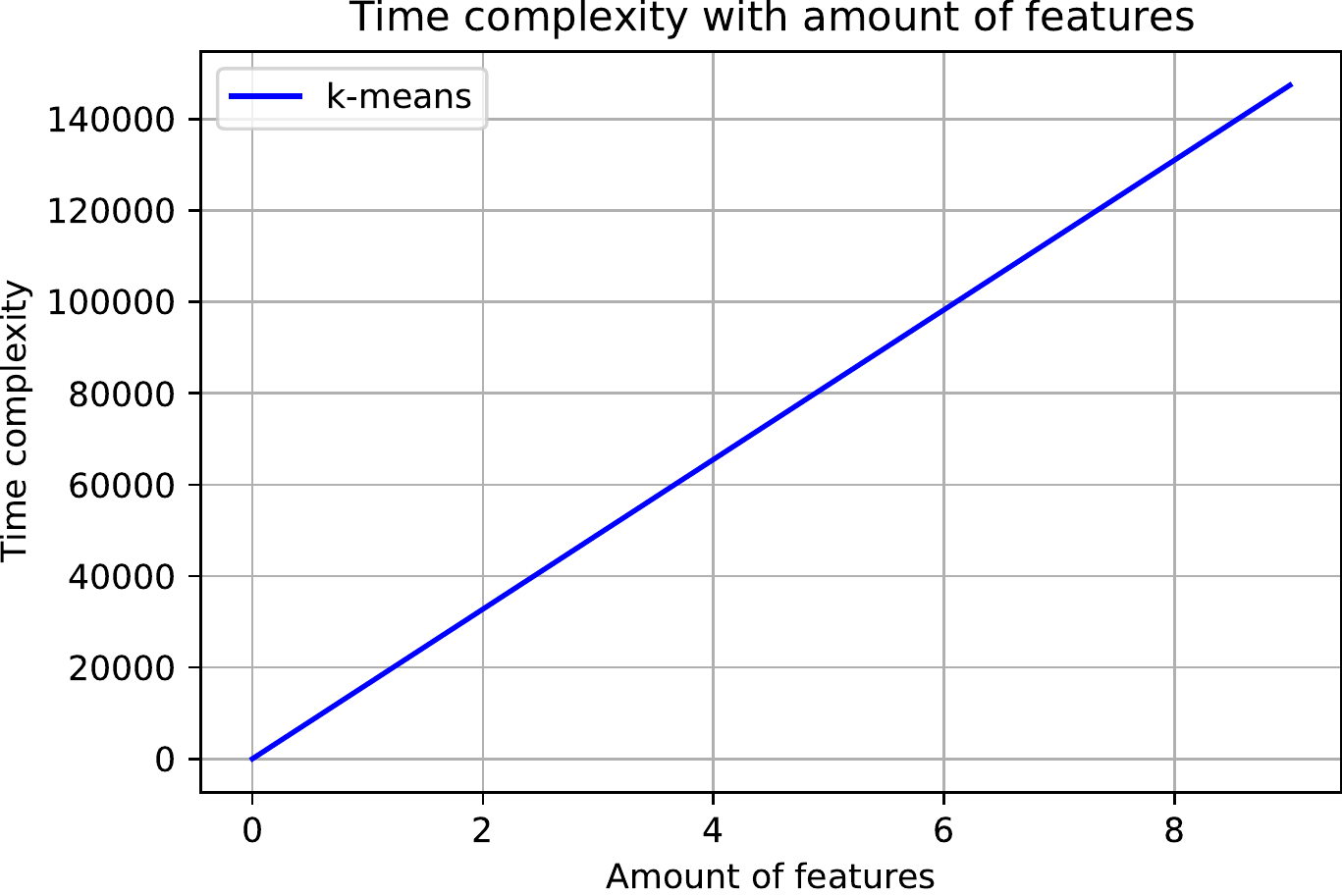}
         \caption{Classical time complexity plot depending on the amount of features.}
         \label{fig:complexity_feature_k}
     \end{subfigure}
     \hfill \\
     \begin{subfigure}[h]{0.45\textwidth}
         \centering
         \includegraphics[width=0.8\textwidth]{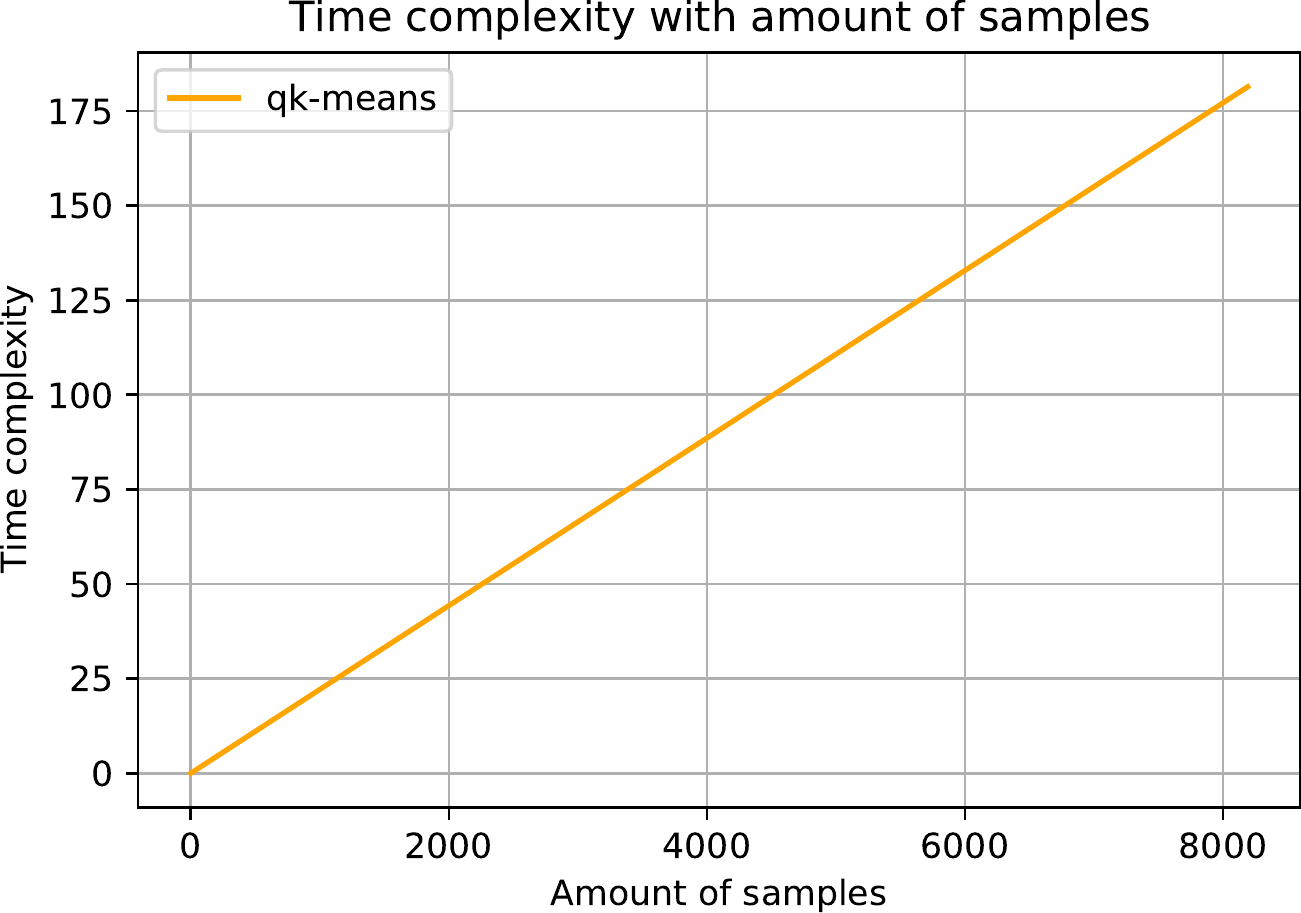}
         \caption{Quantum time complexity plot depending on the amount of samples.}
         \label{fig:complexity_sample_q}
     \end{subfigure}
     \hfill
     \begin{subfigure}[h]{0.45\textwidth}
         \centering
         \includegraphics[width=0.8\textwidth]{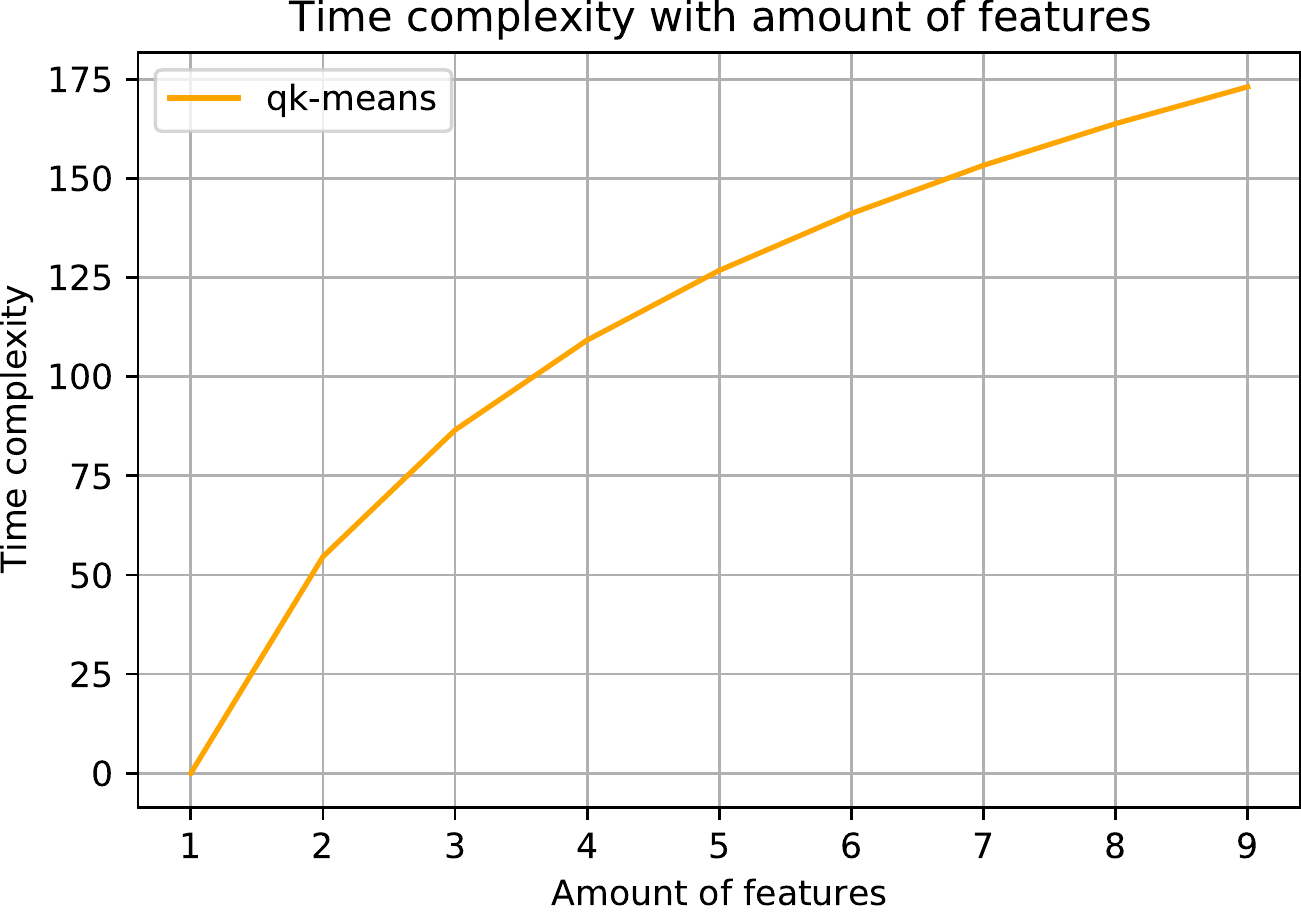}
         \caption{Quantum time complexity plot depending on the amount of features.}
         \label{fig:complexity_feature_q}
     \end{subfigure}
        \caption{Time complexity plots with both theoretical classical and quantum complexity depending on the amount of samples for the plots on the left, and the amount of features for the plots on the right.}
        \label{fig:complexity}
\end{figure*}

Despite one-on-one distance calculations being a slow approach to a KMeans algorithm alternative, our implementation partially makes up for the speed difference by batching the circuits used for distance calculations on IBMQ's quantum devices, since as much as 900 circuits can currently be sent in a single job. This reduces the amount of separate jobs a classical computer has to send to a quantum computer as given \(C = 900\) circuits, \(N\) data points and \(K\) clusters, only \(\frac{NK}{C}\) jobs are sent for each iteration of the algorithm in contrast to the \(NK\) jobs required for each local simulated iteration. In fact, the time complexity of the classical k-means algorithm is \(\mathcal{O}(NKFI)\) where the remaining \(I\) term corresponds to the amount of iterations and the \(F\) term equals the amount of features or dimensions the dataset has, contributed by the time complexity of euclidean distance calculations. The operations a classical computer has to perform to calculate euclidean distances are determined by the expression
$$
D(x,y) = \sqrt{(x_1 - y_1)^2 + (x_2 - y_2)^2 + ... + (x_F - y_F)^2 }
$$
with a small multiple of \(F\) instructions required to calculate one distance. The qk-means accounts for the distance calculation complexity by only requiring measurements on a \(log(F)\) amount of qubits, as after mapping classical data onto qubits, only \(log(F)\) qubits are needed to effectively represent data points with \(F\) features, leaving it with a total time complexity of \(\mathcal{O}(NKlog(F)I/C)\). Figure \ref{fig:complexity} shows the difference in theoretical time complexities found in both algorithms, where a reduction is clear to see. Convergence speed is also accounted for by including qk-means++, a quantum version of the k-means++ initial cluster center guessing strategy in which initial cluster centers are assigned to a data point based on the greatest distance it has to its closest cluster center \cite{10.5555/1283383.1283494}. The general outline of the algorithm is described in Algorithm ~\ref{alg:qkmeans}.

\begin{algorithm}[ht]
\DontPrintSemicolon
\SetAlgoLined
\KwIn{$X, backend, init, n\_clusters, max\_iter, tol$}
\KwOut{$qkmeans$: A trained model}
    $X\longleftarrow \texttt{preprocess}(X)$\;
    $finished\longleftarrow false$\;
    $iteration\longleftarrow 0$\;
    \uIf{$init==$'qk-means++'}{
        $cluster\_centers\longleftarrow \texttt{qkmeans\_plusplus}(X, cluster\_centers, backend)$\;
    }\Else{
        $cluster\_centers\longleftarrow$ sample $n\_clusters$ elements from $X$\;
    }
    \While{not $finished$ and $iteration<max\_iter$}{
        $distances\longleftarrow \texttt{CSWAP}(X, cluster\_centers, backend)$\;
        $labels\longleftarrow$ indices of minimum $distances$\;
        $new\_centroids\longleftarrow$ group $X$ by $labels$ and average\;
        $inertia\longleftarrow \sum\sum|new\_centroids-cluster\_centers|$\;
        \If{$inertia<tol$}{
            $finished\longleftarrow true$\;
        }
        $cluster\_centers\longleftarrow new\_centroids$\;
        $iteration\longleftarrow iteration + 1$\;
    }
 \caption{Quantum K-Means algorithm in pseudocode}
 \label{alg:qkmeans}
\end{algorithm}

\section{Methods}

We start by obtaining IQ signal data from each qubit on a quantum device when applying a ground and a \(\pi\) pulse to obtain both \(\ket{0}\) and \(\ket{1}\) state data. For this, signal data is collected for individual qubits after preparing them in both states. We then utilize the QML algorithm to group clusters together in a similar fashion as would be done with classical ML clustering algorithms, while also performing clustering using classical ML algorithms. The results would be assignment fidelities for each qubit that enable the construction of learning curves with information describing the mean, CI (confidence interval) and convergence rate. We then compare the resulting scores from the classical k-means and the qk-means algorithm on every qubit in order to make conclusions on the performance of both algorithms when discriminating quantum states.

The next set of experiments is oriented towards analyzing the data for information on cross-talk using correlation coefficients and the assignment fidelities as well as the CI of each schedule in a set of schedules. This is done by preparing schedules for the \(\ket{00}\), \(\ket{01}\), \(\ket{10}\) and \(\ket{11}\) states using two neighboring qubits and fitting two ML models on datasets of 1024 data points comprised of the \(\ket{0}\) and \(\ket{1}\) state for a specific qubit (\(\ket{01}\) and \(\ket{00}\) with the least significant bit corresponding to the qubit of choice) with a total of 2048 data points, and states with all four schedules with 4096 data points, respectively. Information on assignment fidelities is then compared to find any significant cross-talk happening on a qubit coupling. We also analyze Pearson coefficients between arrays of signal data retrieved in the form $ r_j(ES_{i,X}, GS_{i,Y}) $ where $ X,Y \in \{I,Q\} $ and qubit $ i $ is in state $ j \in \{0,1\} $ when the other qubit is in the excited state ($ ES $) and ground state ($ GS $). Results from the correlation analysis are contrasted to verify cross-talk on neighboring qubit couples.

\section{Results and discussion}

\subsection{Classical Machine Learning}

When using the k-means clustering algorithm on the \(\ket{00}\), \(\ket{01}\), \(\ket{10}\) and \(\ket{11}\) state schedules for all qubit couplings, we obtained test scores for qubits when its qubit couple is in the ground and excited state, respectively. Tables \ref{table:classical_assignment} and \ref{table:classical_fowlkes} show these scores using assignment fidelity and Fowlkes Mallows score means with confidence intervals. Both scores for inidividual qubits without effects from neighboring qubits are shown in the first column of tables \ref{table:classical_assignment} and \ref{table:classical_fowlkes}.

\begin{table}[]
\begin{tabular}{lll}
   & \multicolumn{1}{c}{Single} & \multicolumn{1}{c}{Both} \\ \hline
Q0 &    0.970 ±0.0054           &       0.973 ±0.0047      \\
Q1 &    0.951 ±0.0123           &       0.949 ±0.0064      \\ \hline
Q1 &    0.949 ±0.0071           &       0.954 ±0.0067      \\
Q2 &    0.976 ±0.0071           &       0.977 ±0.0059      \\ \hline
Q2 &    0.982 ±0.0065           &       0.983 ±0.0056      \\
Q3 &    0.992 ±0.0039           &       0.990 ±0.0032      \\ \hline
Q3 &    0.989 ±0.0037           &       0.988 ±0.0036      \\
Q4 &    0.987 ±0.0056           &       0.988 ±0.0039
\end{tabular}
\caption{Assignment fidelity scores for the test dataset on all neighboring qubit couples with \(\ket{0}\) and \(\ket{1}\) state schedules for single and \(\ket{00}\), \(\ket{01}\), \(\ket{10}\) and \(\ket{11}\) state schedules for both using the classical k-means approach using cross-validation on \(10\) splits.}
\label{table:classical_assignment}
\end{table}

\begin{table}[]
\begin{tabular}{lll}
   & \multicolumn{1}{c}{Single} & \multicolumn{1}{c}{Both} \\ \hline
Q0 &    0.881 ±0.0173           &       0.889 ±0.0175      \\
Q1 &    0.825 ±0.0278           &       0.828 ±0.0255      \\ \hline
Q1 &    0.832 ±0.0435           &       0.821 ±0.0200      \\
Q2 &    0.914 ±0.0294           &       0.911 ±0.0180      \\ \hline
Q2 &    0.926 ±0.0224           &       0.936 ±0.0158      \\
Q3 &    0.969 ±0.0174           &       0.963 ±0.0121      \\ \hline
Q3 &    0.951 ±0.0224           &       0.953 ±0.0133      \\
Q4 &    0.951 ±0.0189           &       0.952 ±0.0134
\end{tabular}
\caption{Fowlkes Mallows scores for the test dataset on all neighboring qubit couples with \(\ket{0}\) and \(\ket{1}\) state schedules for single and \(\ket{00}\), \(\ket{01}\), \(\ket{10}\) and \(\ket{11}\) state schedules for both using the classical k-means approach using cross-validation on \(10\) splits.}
\label{table:classical_fowlkes}
\end{table}

\begin{table}[]
\begin{tabular}{lll}
   & \multicolumn{1}{c}{Single} & \multicolumn{1}{c}{Both} \\ \hline
Q0 &    0.966 ±0.0138           &       0.971 ±0.0008      \\
Q1 &    0.931 ±0.0071           &       0.930 ±0.0165      \\ \hline
Q1 &    0.929 ±0.0037           &       0.935 ±0.0017      \\
Q2 &    0.975 ±0.0009           &       0.974 ±0.0006      \\ \hline
Q2 &    0.975 ±0.0014           &       0.977 ±0.0010      \\
Q3 &    0.987 ±0.0021           &       0.987 ±0.0022      \\ \hline
Q3 &    0.977 ±0.0252           &       0.984 ±0.0010      \\
Q4 &    0.971 ±0.0541           &       0.987 ±0.0001
\end{tabular}
\caption{Assignment fidelity scores for the test dataset on all neighboring qubit couples with \(\ket{0}\) and \(\ket{1}\) state schedules for single and \(\ket{00}\), \(\ket{01}\), \(\ket{10}\) and \(\ket{11}\) state schedules for both using the quantum k-means approach using cross-validation on \(10\) splits.}
\label{table:quantum_assignment}
\end{table}

\begin{table}[]
\begin{tabular}{lll}
   & \multicolumn{1}{c}{Single} & \multicolumn{1}{c}{Both} \\ \hline
Q0 &    0.875 ±0.0439           &       0.890 ±0.0029      \\
Q1 &    0.762 ±0.0210           &       0.762 ±0.0424      \\ \hline
Q1 &    0.756 ±0.0211           &       0.753 ±0.0263      \\
Q2 &    0.906 ±0.0029           &       0.905 ±0.0090      \\ \hline
Q2 &    0.907 ±0.0048           &       0.900 ±0.0939      \\
Q3 &    0.915 ±0.0036           &       0.951 ±0.0093      \\ \hline
Q3 &    0.937 ±0.0098           &       0.896 ±0.1075      \\
Q4 &    0.837 ±0.1470           &       0.889 ±0.1348
\end{tabular}
\caption{Fowlkes Mallows scores for the test dataset on all neighboring qubit couples with \(\ket{0}\) and \(\ket{1}\) state schedules for single and \(\ket{00}\), \(\ket{01}\), \(\ket{10}\) and \(\ket{11}\) state schedules for both using the quantum k-means approach using cross-validation on \(10\) splits.}
\label{table:quantum_fowlkes}
\end{table}

\begin{table}[]
\begin{tabular}{lllll}
   & (0, 1)     & (1, 2)     & (2, 3)     & (3, 4)     \\ \hline
$ r_0(ES_{i,X}, GS_{i,Y}) $     & -0.0236 & 0.0070 & -0.0265 & -0.0058 \\ \hline
$ r_0(ES_{i,Y}, GS_{i,X}) $     & -0.0161 & 0.0063 & -0.0057 & 0.0042  \\ \hline
$ r_1(ES_{i,X}, GS_{i,Y}) $     & 0.0017  & 0.0011 & 0.1814  & -0.0434 \\ \hline
$ r_1(ES_{i,Y}, GS_{i,X}) $     & -0.0211 & 0.0214 & 0.2147  & 0.0166  \\ \hline
$ r_0(ES_{i+1,X}, GS_{i+1,Y}) $ & 0.0311  & 0.0869 & 0.0224  & 0.0210  \\ \hline
$ r_0(ES_{i+1,Y}, GS_{i+1,X}) $ & -0.0394 & 0.1927 & 0.0369  & -0.0102 \\ \hline
$ r_1(ES_{i+1,X}, GS_{i+1,Y}) $ & 0.02562 & 0.1923 & 0.0227  & -0.0086 \\ \hline
$ r_1(ES_{i+1,Y}, GS_{i+1,X}) $ & -0.0012 & 0.1111 & -0.0111 & 0.0033
\end{tabular}
\caption{Pearson correlation coefficients in the form $ r_j(ES_{i,X}, GS_{i,Y}) $ where $ X,Y \in \{I,Q\} $ and qubit $ i $ is in state $ j \in \{0,1\} $ when the other qubit is in the excited state and ground state and $ i $ is the first qubit of the couple, for each qubit coupling on the device.}
\label{table:pearson_correlation}
\end{table}

\subsection{Quantum Machine Learning}

\begin{figure}
\centering
\includegraphics[width=0.5\textwidth]{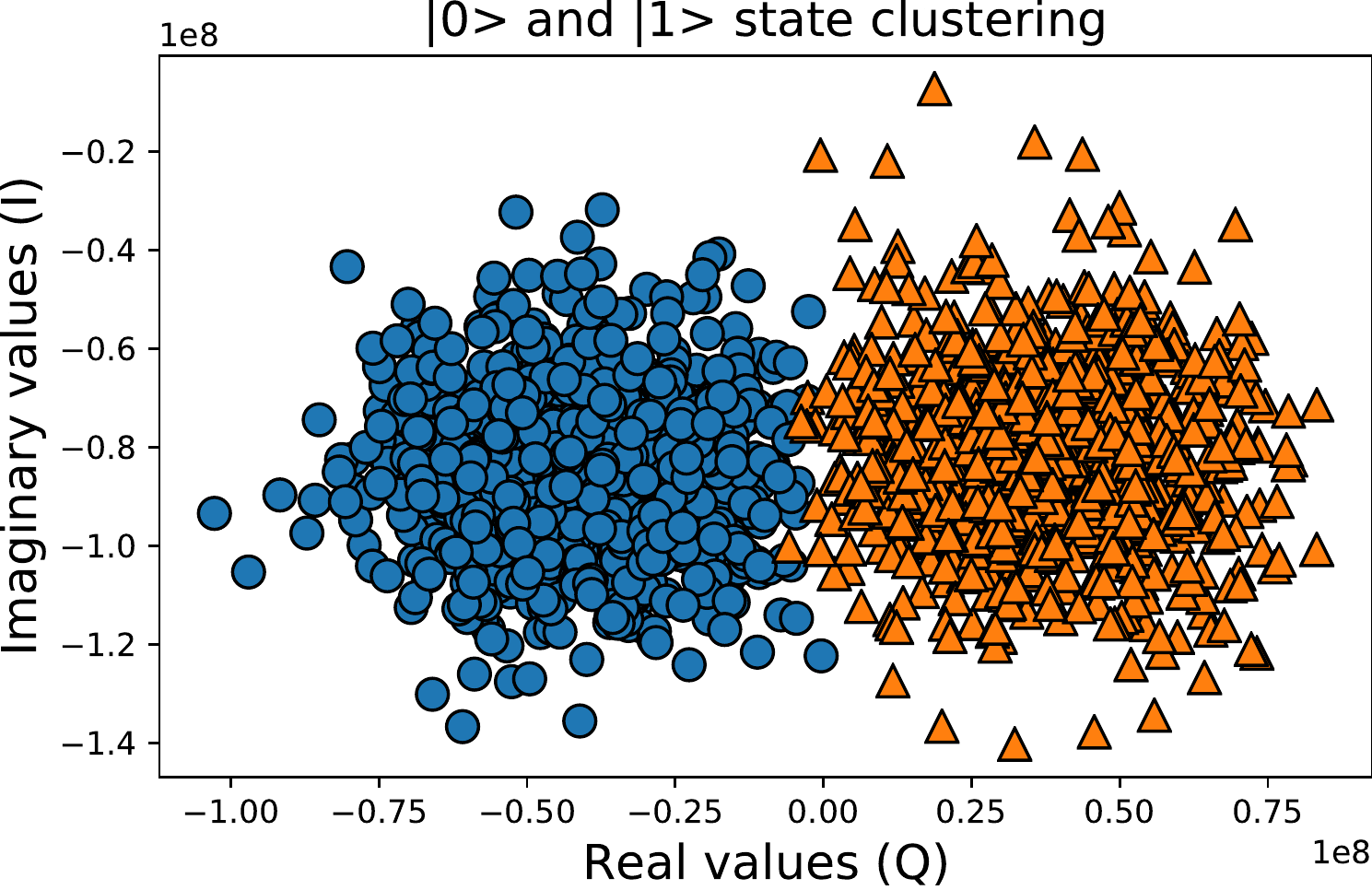}
\caption{Result of running the qk-means clustering algorithm on the ibmq\_qasm simulator for signal data retrieved from qubit 0 of ibmq\_bogota when applying schedules \(\ket{00}\), \(\ket{01}\), \(\ket{10}\) and \(\ket{11}\).}
\label{fig:clustering}
\end{figure}

Results for the QML approach are obtained in a similar manner as the classical ML results but with the qk-means as the selected clustering algorithm. Results for neighboring qubit couples are shown in tables \ref{table:quantum_assignment} and \ref{table:quantum_fowlkes} while results for inidividual qubits are shown in the first column of tables \ref{table:quantum_assignment} and \ref{table:quantum_fowlkes}. Clustering of signal data on qubit 0 is shown in Figure \ref{fig:clustering}.

\subsection{Pearson correlation coefficients}

\begin{figure}
\centering
\includegraphics[width=0.5\textwidth]{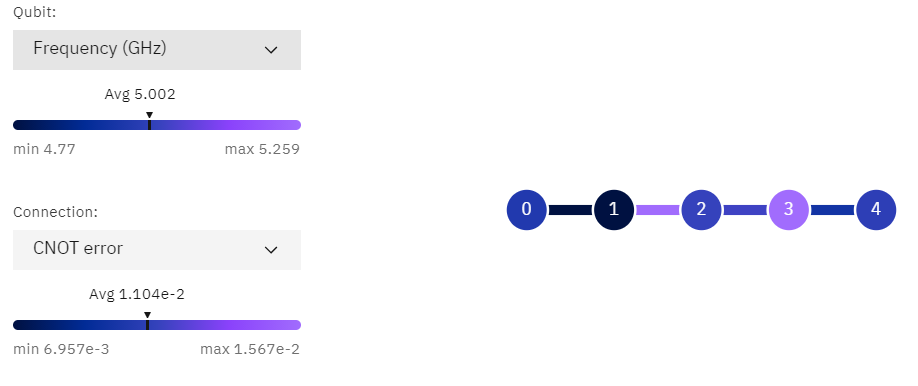}
\caption{The qubit coupling map of the ibmq\_bogota device at the time of signal data retrieval. The map also shows the frequency values of each qubit and the CNOT errors between each pair of connected qubits,}
\label{fig:coupling_map}
\end{figure}

\begin{figure*}
     \centering
     \begin{subfigure}[h]{0.45\textwidth}
         \centering
         \includegraphics[width=0.8\textwidth]{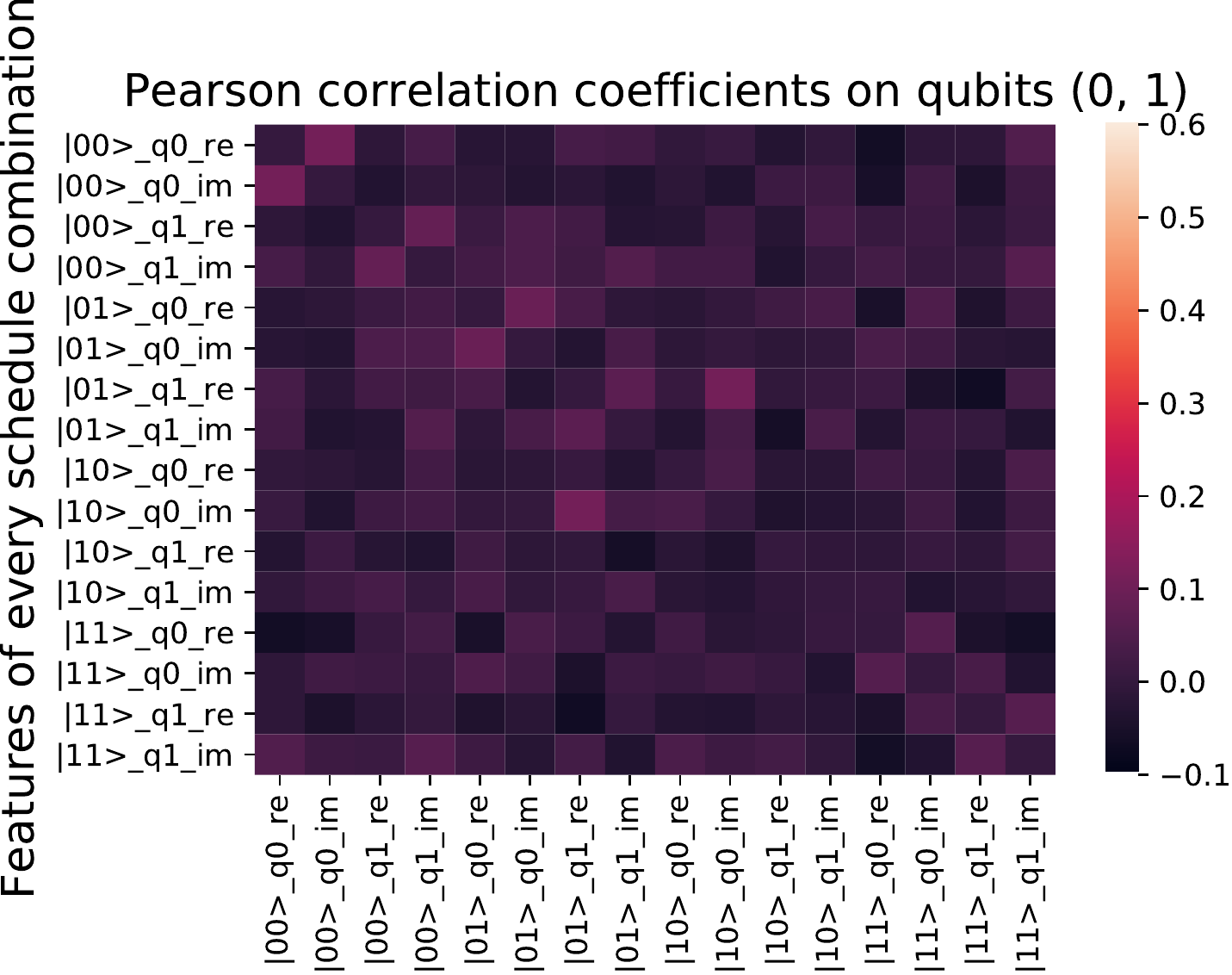}
         \caption{Qubit coupling (0, 1).}
         \label{fig:heatmap_0_1}
     \end{subfigure}
     \hfill
     \begin{subfigure}[h]{0.45\textwidth}
         \centering
         \includegraphics[width=0.8\textwidth]{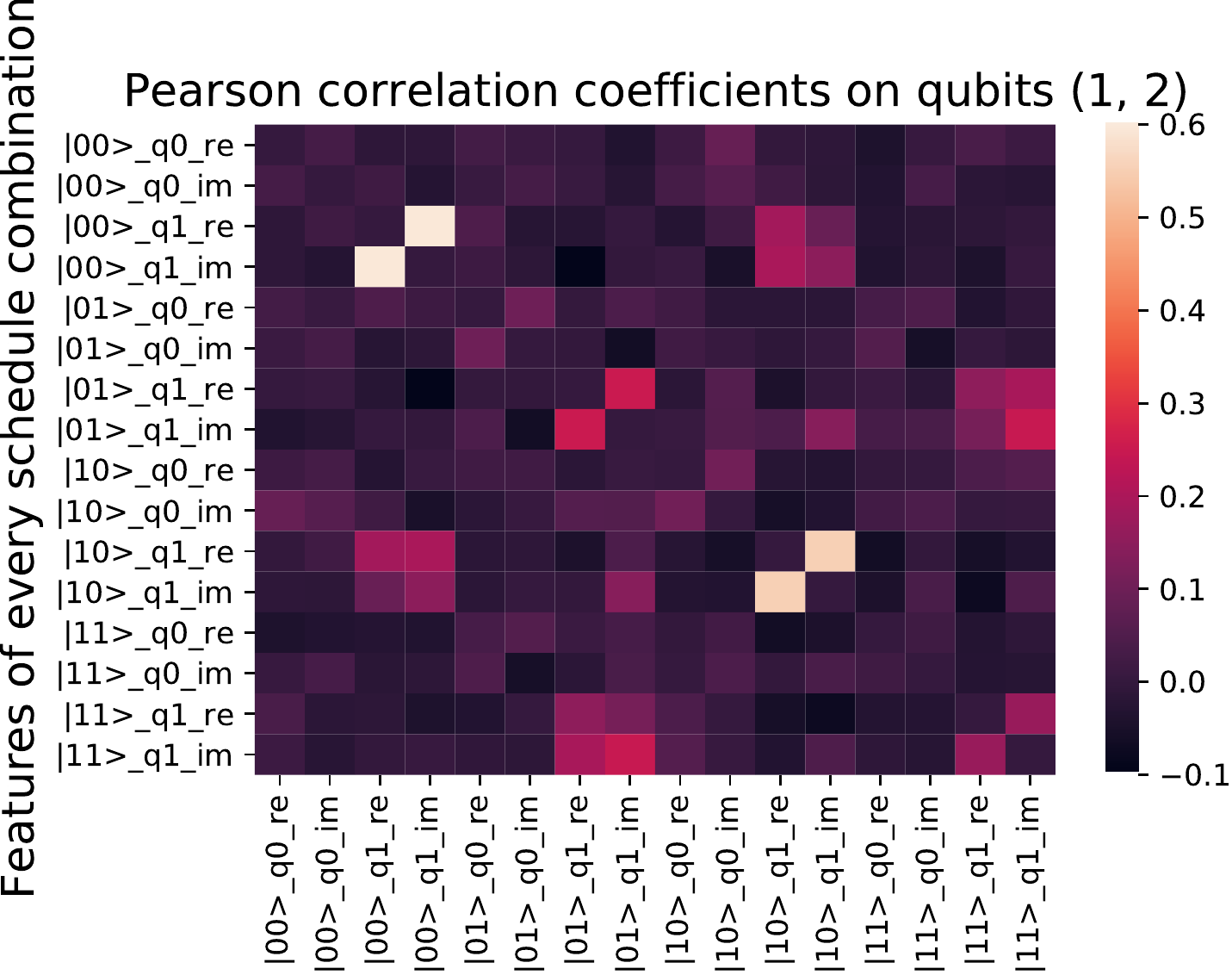}
         \caption{Qubit coupling (1, 2).}
         \label{fig:heatmap_1_2}
     \end{subfigure}
     \hfill \\
     \begin{subfigure}[h]{0.45\textwidth}
         \centering
         \includegraphics[width=0.8\textwidth]{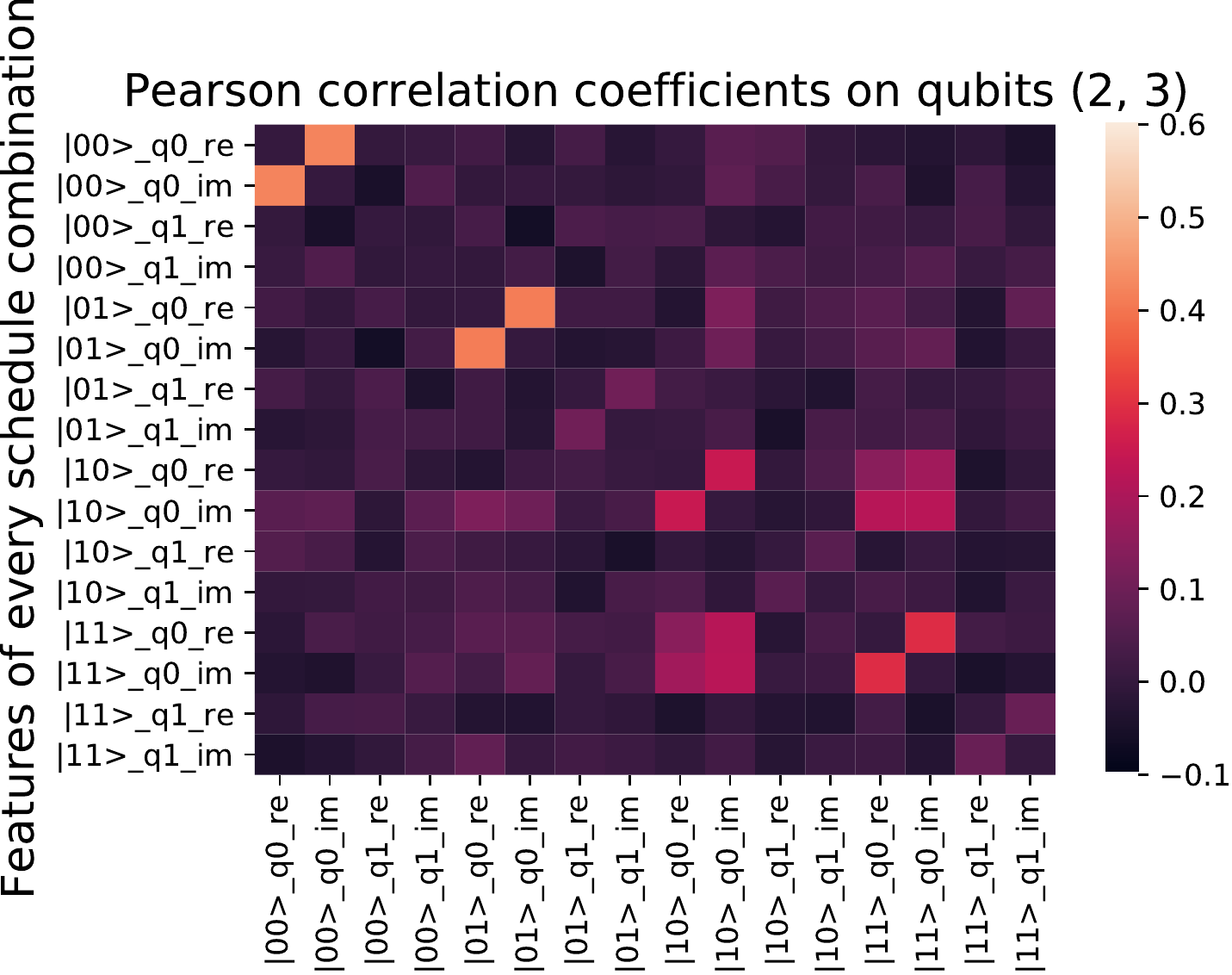}
         \caption{Qubit coupling (2, 3).}
         \label{fig:heatmap_2_3}
     \end{subfigure}
     \hfill
     \begin{subfigure}[h]{0.45\textwidth}
         \centering
         \includegraphics[width=0.8\textwidth]{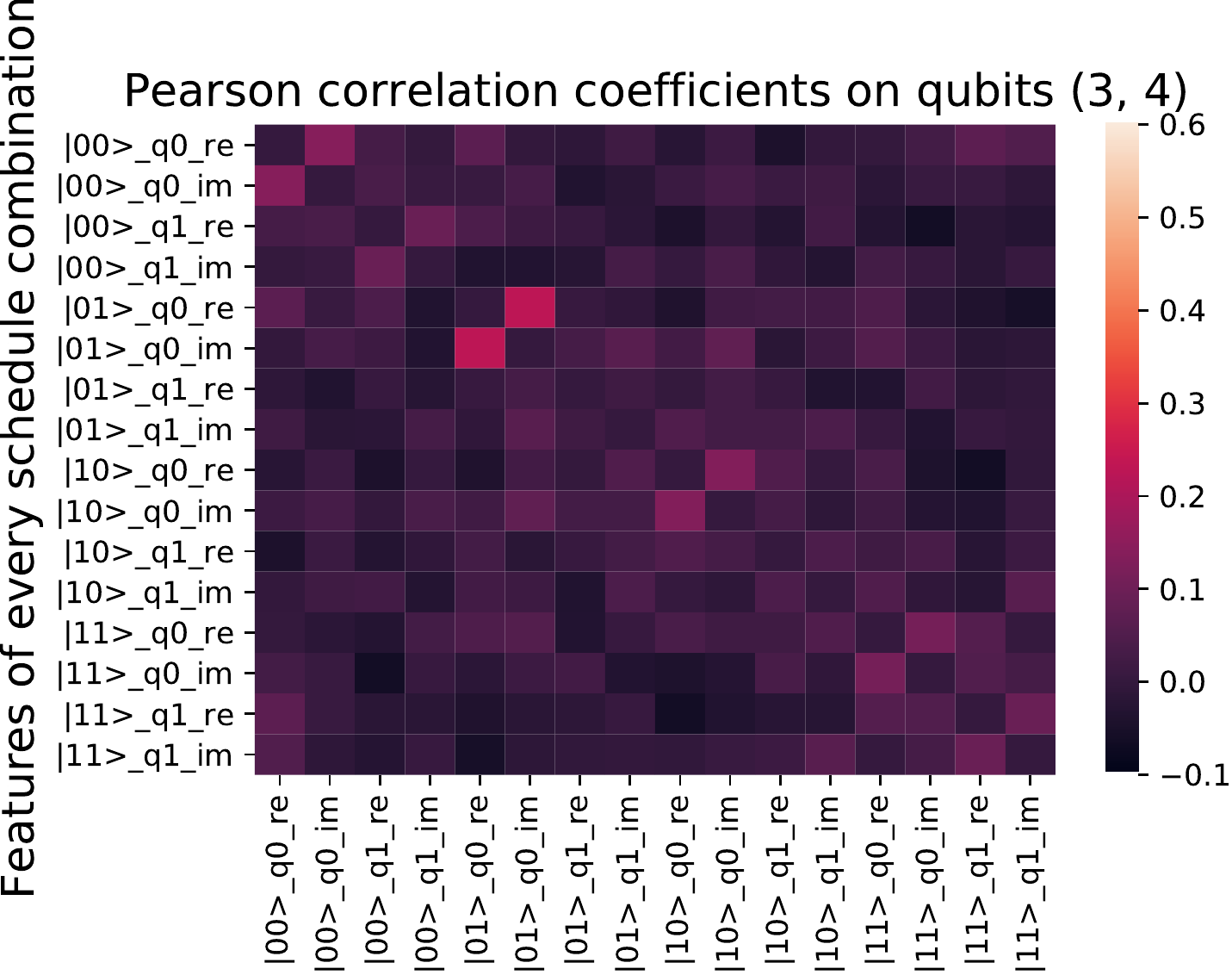}
         \caption{Qubit coupling (3, 4).}
         \label{fig:heatmap_3_4}
     \end{subfigure}
        \caption{Heatmaps showing the Pearson correlation coefficients between each array belonging to the signal data of the four prepared schedules acting on each qubit coupling in the ibmq\_bogota device. The axes labels follow the following format for the signal data received: \{state\}\_\{couple qubit\}\_\{real or imaginary part of the complex number\}.}
        \label{fig:heatmaps_coupling}
\end{figure*}

With the IQ signal data retrieved on all four schedule variations, we extract arrays of each feature for a total of 8 arrays. These arrays are used as input to analyze the Pearson correlation coefficients between each pair of arrays. This procedure is implemented for each qubit coupling on the ibmq\_bogota device with the mapping shown in Figure~\ref{fig:coupling_map}. A heatmap that describes the full spectrum of correlations between neighboring qubits is illustrated in Figure~\ref{fig:heatmaps_coupling}. The Pearson correlation coefficients of interest in the form $ r_j(ES_{i,X}, GS_{i,Y}) $ for each qubit coupling are shown in table \ref{table:pearson_correlation}. The further the coefficients are from zero, the higher the correlation.

Similar test and training scores were observed when using the k-means and the qk-means algorithms on single qubits when no other qubit was in the excited state. The assignment fidelity of the qk-means was slightly lower than that of the k-means for cases when the clusters weren't visually separable while the k-means held better Fowlkes Mallows scores. This result describes how well both algorithms cluster the presented data, as a not so large difference in assignment fidelity and Fowlkes Mallows scores can be found. When observing the Pearson correlation heatmaps, the most significant correlation can be seen between qubits 1 and 2 as off-diagonal coefficients are the highest, followed by qubit couples (2, 3), (3, 4), and (0, 1) in descending order, with qubit couple (0, 1) presenting the lowest scale in Pearson correlation coefficients. This is also verified with the specific correlations in the form $ r_j(ES_{i,X}, GS_{i,Y}) $, where weak correlation can be found in qubit couples (1, 2) and (2, 3). We use the classical k-means assignment fidelity to perform the correlation analysis due to it presenting the most stable results out of both clustering algorithm and scores. Weak correlation in qubit couples (1, 2) and (2, 3) is corroborated with the ML scores for cases "single" and "both", showing scores that are further apart than the same cases on the rest of the qubit couplings. The greatest difference between scores on "single" and "both" cases was found on qubit couple (1, 2) with a 5\% difference on qubit 1.

\section{Conclusions}
We implemented a quantum k-means variant with the Qiskit library compatible with functions from the scikit-learn library. The training and testing scores of this algorithm were at par with the k-means algorithm except when clusters weren't visually separable, in which case only a small difference in scores was observed. We demonstrated the use of a cross-talk pulse-level benchmarking scheme on IBM's ibmq\_bogota quantum computer through the use of ML, QML and Pearson correlation coefficients, finding enough evidence to determine weak correlation in qubit couples (1, 2) and (2, 3). After analyzing the training scores, we also conclude that the 1 qubit has the worst performance at readout evidenced by the signal data not being visually separable and the low scores obtained on both clustering algorithms compared to the other qubits. Its poor performance is further verified by the calibration data showing a high readout error of 8,4\%. Training times for qk-means were similar to the training times on a k-means implementation that was not optimized, whereas the k-means algorithm available on the scikit-learn library was faster. Future work for this project is an optimization of the qk-means algorithm for faster training times given the current constraints of universal quantum computers like the one used for this analysis. Faster training times would allow to collect more data on assignment fidelities after training and to keep quantum device discriminator calibration as a quick process. To that effect, qk-means++ and a way to calculate more distances simultaneously has been implemented and other strategies are being considered to achieve faster training times. Despite there being the possibility to batch several circuits at a single time on a quantum computer for the qk-means, the main difference in speed the scikit-learn implementation of k-means has against the qk-means comes from the replacement of one-on-one distance measurements for distance measurements using a small amount of dot products, greatly reducing its training time on a classical computer.

\section*{Acknowledgment}

This work was supported in part by the U.S. Department of Energy, Office of Science, Office of Workforce Development for Teachers and Scientists (WDTS) under the Science Undergraduate Laboratory Internship program. This work was completed through Oak Ridge National Laboratory with the collaboration of staff scientists Prasanna Date in the Beyond Moore group and Raphael C. Pooser in the Quantum Information Sciences group. Special thanks to professor Javier F. Botia from Universidad de Antioquia for useful conversations on the algorithm.

\bibliographystyle{IEEEtran}
\bibliography{references}

\begin{thebibliography}{10}
\providecommand{\url}[1]{#1}
\csname url@samestyle\endcsname
\providecommand{\newblock}{\relax}
\providecommand{\bibinfo}[2]{#2}
\providecommand{\BIBentrySTDinterwordspacing}{\spaceskip=0pt\relax}
\providecommand{\BIBentryALTinterwordstretchfactor}{4}
\providecommand{\BIBentryALTinterwordspacing}{\spaceskip=\fontdimen2\font plus
\BIBentryALTinterwordstretchfactor\fontdimen3\font minus
  \fontdimen4\font\relax}
\providecommand{\BIBforeignlanguage}[2]{{%
\expandafter\ifx\csname l@#1\endcsname\relax
\typeout{** WARNING: IEEEtran.bst: No hyphenation pattern has been}%
\typeout{** loaded for the language `#1'. Using the pattern for}%
\typeout{** the default language instead.}%
\else
\language=\csname l@#1\endcsname
\fi
#2}}
\providecommand{\BIBdecl}{\relax}
\BIBdecl

\bibitem{Biamonte2017}
\BIBentryALTinterwordspacing
J.~Biamonte, P.~Wittek, N.~Pancotti, P.~Rebentrost, N.~Wiebe, and S.~Lloyd,
  ``Quantum machine learning,'' \emph{Nature}, vol. 549, no. 7671, pp.
  195--202, Sep 2017. [Online]. Available:
  \url{https://doi.org/10.1038/nature23474}
\BIBentrySTDinterwordspacing

\bibitem{PhysRevLett.103.150502}
\BIBentryALTinterwordspacing
A.~W. Harrow, A.~Hassidim, and S.~Lloyd, ``Quantum algorithm for linear systems
  of equations,'' \emph{Phys. Rev. Lett.}, vol. 103, p. 150502, Oct 2009.
  [Online]. Available:
  \url{https://link.aps.org/doi/10.1103/PhysRevLett.103.150502}
\BIBentrySTDinterwordspacing

\bibitem{PhysRevLett.109.050505}
\BIBentryALTinterwordspacing
N.~Wiebe, D.~Braun, and S.~Lloyd, ``Quantum algorithm for data fitting,''
  \emph{Phys. Rev. Lett.}, vol. 109, p. 050505, Aug 2012. [Online]. Available:
  \url{https://link.aps.org/doi/10.1103/PhysRevLett.109.050505}
\BIBentrySTDinterwordspacing

\bibitem{Childs_2017}
\BIBentryALTinterwordspacing
A.~M. Childs, R.~Kothari, and R.~D. Somma, ``Quantum algorithm for systems of
  linear equations with exponentially improved dependence on precision,''
  \emph{SIAM Journal on Computing}, vol.~46, no.~6, p. 1920–1950, Jan 2017.
  [Online]. Available: \url{http://dx.doi.org/10.1137/16M1087072}
\BIBentrySTDinterwordspacing

\bibitem{Preskill2018quantumcomputingin}
\BIBentryALTinterwordspacing
J.~Preskill, ``Quantum {C}omputing in the {NISQ} era and beyond,''
  \emph{{Quantum}}, vol.~2, p.~79, Aug. 2018. [Online]. Available:
  \url{https://doi.org/10.22331/q-2018-08-06-79}
\BIBentrySTDinterwordspacing

\bibitem{Feng_2016}
\BIBentryALTinterwordspacing
G.~Feng, J.~J. Wallman, B.~Buonacorsi, F.~H. Cho, D.~K. Park, T.~Xin, D.~Lu,
  J.~Baugh, and R.~Laflamme, ``Estimating the coherence of noise in quantum
  control of a solid-state qubit,'' \emph{Physical Review Letters}, vol. 117,
  no.~26, Dec 2016. [Online]. Available:
  \url{http://dx.doi.org/10.1103/PhysRevLett.117.260501}
\BIBentrySTDinterwordspacing

\bibitem{PhysRevResearch.2.043418}
\BIBentryALTinterwordspacing
Y.~He, L.~Ji, Y.~Wang, L.~Qiu, J.~Zhao, Y.~Ma, X.~Huang, S.~Wu, and D.~E.
  Chang, ``Atomic spin-wave control and spin-dependent kicks with shaped
  subnanosecond pulses,'' \emph{Phys. Rev. Research}, vol.~2, p. 043418, Dec
  2020. [Online]. Available:
  \url{https://link.aps.org/doi/10.1103/PhysRevResearch.2.043418}
\BIBentrySTDinterwordspacing

\bibitem{Nachman2020}
\BIBentryALTinterwordspacing
B.~Nachman, M.~Urbanek, W.~A. de~Jong, and C.~W. Bauer, ``Unfolding quantum
  computer readout noise,'' \emph{npj Quantum Information}, vol.~6, no.~1,
  p.~84, Sep 2020. [Online]. Available:
  \url{https://doi.org/10.1038/s41534-020-00309-7}
\BIBentrySTDinterwordspacing

\bibitem{10.1145/3373376.3378477}
\BIBentryALTinterwordspacing
P.~Murali, D.~C. Mckay, M.~Martonosi, and A.~Javadi-Abhari, ``Software
  mitigation of crosstalk on noisy intermediate-scale quantum computers,'' in
  \emph{Proceedings of the Twenty-Fifth International Conference on
  Architectural Support for Programming Languages and Operating Systems}, ser.
  ASPLOS '20.\hskip 1em plus 0.5em minus 0.4em\relax New York, NY, USA:
  Association for Computing Machinery, 2020, p. 1001–1016. [Online].
  Available: \url{https://doi.org/10.1145/3373376.3378477}
\BIBentrySTDinterwordspacing

\bibitem{8614500}
M.~Brink, J.~M. Chow, J.~Hertzberg, E.~Magesan, and S.~Rosenblatt, ``Device
  challenges for near term superconducting quantum processors: frequency
  collisions,'' in \emph{2018 IEEE International Electron Devices Meeting
  (IEDM)}, 2018, pp. 6.1.1--6.1.3.

\bibitem{Sarovar2020detectingcrosstalk}
\BIBentryALTinterwordspacing
M.~Sarovar, T.~Proctor, K.~Rudinger, K.~Young, E.~Nielsen, and R.~Blume-Kohout,
  ``Detecting crosstalk errors in quantum information processors,''
  \emph{{Quantum}}, vol.~4, p. 321, Sep. 2020. [Online]. Available:
  \url{https://doi.org/10.22331/q-2020-09-11-321}
\BIBentrySTDinterwordspacing

\bibitem{HU_2002}
\BIBentryALTinterwordspacing
X.~HU, R.~DE~SOUSA, and S.~D. SARMA, ``Decoherence and dephasing in spin-based
  solid state quantum computers,'' \emph{Foundations of Quantum Mechanics in
  the Light of New Technology}, Oct 2002. [Online]. Available:
  \url{http://dx.doi.org/10.1142/9789812776716\_0001}
\BIBentrySTDinterwordspacing

\bibitem{werninghaus2020leakage}
M.~Werninghaus, D.~J. Egger, F.~Roy, S.~Machnes, F.~K. Wilhelm, and S.~Filipp,
  ``Leakage reduction in fast superconducting qubit gates via optimal
  control,'' 2020.

\bibitem{PhysRevA.93.060302}
\BIBentryALTinterwordspacing
S.~Sheldon, E.~Magesan, J.~M. Chow, and J.~M. Gambetta, ``Procedure for
  systematically tuning up cross-talk in the cross-resonance gate,''
  \emph{Phys. Rev. A}, vol.~93, p. 060302, Jun 2016. [Online]. Available:
  \url{https://link.aps.org/doi/10.1103/PhysRevA.93.060302}
\BIBentrySTDinterwordspacing

\bibitem{Chasseur_2015}
\BIBentryALTinterwordspacing
T.~Chasseur and F.~K. Wilhelm, ``Complete randomized benchmarking protocol
  accounting for leakage errors,'' \emph{Physical Review A}, vol.~92, no.~4,
  Oct 2015. [Online]. Available:
  \url{http://dx.doi.org/10.1103/PhysRevA.92.042333}
\BIBentrySTDinterwordspacing

\bibitem{Knill_2008}
\BIBentryALTinterwordspacing
E.~Knill, D.~Leibfried, R.~Reichle, J.~Britton, R.~B. Blakestad, J.~D. Jost,
  C.~Langer, R.~Ozeri, S.~Seidelin, and D.~J. Wineland, ``Randomized
  benchmarking of quantum gates,'' \emph{Physical Review A}, vol.~77, no.~1,
  Jan 2008. [Online]. Available:
  \url{http://dx.doi.org/10.1103/PhysRevA.77.012307}
\BIBentrySTDinterwordspacing

\bibitem{PhysRevA.101.042308}
\BIBentryALTinterwordspacing
J.~W.~O. Garmon, R.~C. Pooser, and E.~F. Dumitrescu, ``Benchmarking noise
  extrapolation with the openpulse control framework,'' \emph{Phys. Rev. A},
  vol. 101, p. 042308, Apr 2020. [Online]. Available:
  \url{https://link.aps.org/doi/10.1103/PhysRevA.101.042308}
\BIBentrySTDinterwordspacing

\bibitem{10.5555/1972505}
M.~A. Nielsen and I.~L. Chuang, \emph{Quantum Computation and Quantum
  Information: 10th Anniversary Edition}, 10th~ed.\hskip 1em plus 0.5em minus
  0.4em\relax USA: Cambridge University Press, 2011.

\bibitem{Magesan_2015}
\BIBentryALTinterwordspacing
E.~Magesan, J.~M. Gambetta, A.~Córcoles, and J.~M. Chow, ``Machine learning
  for discriminating quantum measurement trajectories and improving readout,''
  \emph{Physical Review Letters}, vol. 114, no.~20, May 2015. [Online].
  Available: \url{http://dx.doi.org/10.1103/PhysRevLett.114.200501}
\BIBentrySTDinterwordspacing

\bibitem{Alexander_2020}
\BIBentryALTinterwordspacing
T.~Alexander, N.~Kanazawa, D.~J. Egger, L.~Capelluto, C.~J. Wood,
  A.~Javadi-Abhari, and D.~C~McKay, ``Qiskit pulse: programming quantum
  computers through the cloud with pulses,'' \emph{Quantum Science and
  Technology}, vol.~5, no.~4, p. 044006, Aug 2020. [Online]. Available:
  \url{http://dx.doi.org/10.1088/2058-9565/aba404}
\BIBentrySTDinterwordspacing

\bibitem{mckay2018qiskit}
D.~C. McKay, T.~Alexander, L.~Bello, M.~J. Biercuk, L.~Bishop, J.~Chen, J.~M.
  Chow, A.~D. Córcoles, D.~Egger, S.~Filipp, J.~Gomez, M.~Hush,
  A.~Javadi-Abhari, D.~Moreda, P.~Nation, B.~Paulovicks, E.~Winston, C.~J.
  Wood, J.~Wootton, and J.~M. Gambetta, ``Qiskit backend specifications for
  openqasm and openpulse experiments,'' 2018.

\bibitem{doi:10.1063/1.4813269}
\BIBentryALTinterwordspacing
J.~B. Chang, M.~R. Vissers, A.~D. Córcoles, M.~Sandberg, J.~Gao, D.~W.
  Abraham, J.~M. Chow, J.~M. Gambetta, M.~Beth~Rothwell, G.~A. Keefe,
  M.~Steffen, and D.~P. Pappas, ``Improved superconducting qubit coherence
  using titanium nitride,'' \emph{Applied Physics Letters}, vol. 103, no.~1, p.
  012602, 2013. [Online]. Available: \url{https://doi.org/10.1063/1.4813269}
\BIBentrySTDinterwordspacing

\bibitem{sklearn_api}
L.~Buitinck, G.~Louppe, M.~Blondel, F.~Pedregosa, A.~Mueller, O.~Grisel,
  V.~Niculae, P.~Prettenhofer, A.~Gramfort, J.~Grobler, R.~Layton,
  J.~VanderPlas, A.~Joly, B.~Holt, and G.~Varoquaux, ``{API} design for machine
  learning software: experiences from the scikit-learn project,'' in \emph{ECML
  PKDD Workshop: Languages for Data Mining and Machine Learning}, 2013, pp.
  108--122.

\bibitem{Schuld_2014}
\BIBentryALTinterwordspacing
M.~Schuld, I.~Sinayskiy, and F.~Petruccione, ``An introduction to quantum
  machine learning,'' \emph{Contemporary Physics}, vol.~56, no.~2, p.
  172–185, Oct 2014. [Online]. Available:
  \url{http://dx.doi.org/10.1080/00107514.2014.964942}
\BIBentrySTDinterwordspacing

\bibitem{adachi2015application}
S.~H. Adachi and M.~P. Henderson, ``Application of quantum annealing to
  training of deep neural networks,'' 2015.

\bibitem{Brassard_2002}
\BIBentryALTinterwordspacing
G.~Brassard, P.~Høyer, M.~Mosca, and A.~Tapp, ``Quantum amplitude
  amplification and estimation,'' \emph{Quantum Computation and Information},
  p. 53–74, 2002. [Online]. Available:
  \url{http://dx.doi.org/10.1090/conm/305/05215}
\BIBentrySTDinterwordspacing

\bibitem{PhysRevA.89.062315}
\BIBentryALTinterwordspacing
G.~H. Low, T.~J. Yoder, and I.~L. Chuang, ``Quantum inference on bayesian
  networks,'' \emph{Phys. Rev. A}, vol.~89, p. 062315, Jun 2014. [Online].
  Available: \url{https://link.aps.org/doi/10.1103/PhysRevA.89.062315}
\BIBentrySTDinterwordspacing

\bibitem{Lloyd2014}
\BIBentryALTinterwordspacing
S.~Lloyd, M.~Mohseni, and P.~Rebentrost, ``Quantum principal component
  analysis,'' \emph{Nature Physics}, vol.~10, no.~9, pp. 631--633, Sep 2014.
  [Online]. Available: \url{https://doi.org/10.1038/nphys3029}
\BIBentrySTDinterwordspacing

\bibitem{PhysRevLett.113.130503}
\BIBentryALTinterwordspacing
P.~Rebentrost, M.~Mohseni, and S.~Lloyd, ``Quantum support vector machine for
  big data classification,'' \emph{Phys. Rev. Lett.}, vol. 113, p. 130503, Sep
  2014. [Online]. Available:
  \url{https://link.aps.org/doi/10.1103/PhysRevLett.113.130503}
\BIBentrySTDinterwordspacing

\bibitem{wiebe2016quantum}
N.~Wiebe, A.~Kapoor, and K.~M. Svore, ``Quantum perceptron models,'' 2016.

\bibitem{PhysRevLett.117.130501}
\BIBentryALTinterwordspacing
V.~Dunjko, J.~M. Taylor, and H.~J. Briegel, ``Quantum-enhanced machine
  learning,'' \emph{Phys. Rev. Lett.}, vol. 117, p. 130501, Sep 2016. [Online].
  Available: \url{https://link.aps.org/doi/10.1103/PhysRevLett.117.130501}
\BIBentrySTDinterwordspacing

\bibitem{Date2021}
\BIBentryALTinterwordspacing
P.~Date, D.~Arthur, and L.~Pusey-Nazzaro, ``Qubo formulations for training
  machine learning models,'' \emph{Scientific Reports}, vol.~11, no.~1, p.
  10029, May 2021. [Online]. Available:
  \url{https://doi.org/10.1038/s41598-021-89461-4}
\BIBentrySTDinterwordspacing

\bibitem{arthur2020balanced}
D.~Arthur and P.~Date, ``Balanced k-means clustering on an adiabatic quantum
  computer,'' 2020.

\bibitem{Aaronson2015}
\BIBentryALTinterwordspacing
S.~Aaronson, ``Read the fine print,'' \emph{Nature Physics}, vol.~11, no.~4,
  pp. 291--293, Apr 2015. [Online]. Available:
  \url{https://doi.org/10.1038/nphys3272}
\BIBentrySTDinterwordspacing

\bibitem{Peruzzo2014}
\BIBentryALTinterwordspacing
A.~Peruzzo, J.~McClean, P.~Shadbolt, M.-H. Yung, X.-Q. Zhou, P.~J. Love,
  A.~Aspuru-Guzik, and J.~L. O'Brien, ``A variational eigenvalue solver on a
  photonic quantum processor,'' \emph{Nature Communications}, vol.~5, no.~1, p.
  4213, Jul 2014. [Online]. Available: \url{https://doi.org/10.1038/ncomms5213}
\BIBentrySTDinterwordspacing

\bibitem{scikit-learn}
F.~Pedregosa, G.~Varoquaux, A.~Gramfort, V.~Michel, B.~Thirion, O.~Grisel,
  M.~Blondel, P.~Prettenhofer, R.~Weiss, V.~Dubourg, J.~Vanderplas, A.~Passos,
  D.~Cournapeau, M.~Brucher, M.~Perrot, and E.~Duchesnay, ``Scikit-learn:
  Machine learning in {P}ython,'' \emph{Journal of Machine Learning Research},
  vol.~12, pp. 2825--2830, 2011.

\bibitem{10.5555/1283383.1283494}
D.~Arthur and S.~Vassilvitskii, ``K-means++: The advantages of careful
  seeding,'' in \emph{Proceedings of the Eighteenth Annual ACM-SIAM Symposium
  on Discrete Algorithms}, ser. SODA '07.\hskip 1em plus 0.5em minus
  0.4em\relax USA: Society for Industrial and Applied Mathematics, 2007, p.
  1027–1035.

\end{thebibliography}

\end{document}